# Atomic-Scale Quantum Control of Single Spin Defects in a Two-Dimensional Semiconductor

Kwan Ho Au-Yeung[1,2,†], Wantong Huang[1,†], Johanna Matusche[1,†], Paul Greule[1], Jonas Arnold[1], Lovis Hardeweg[1], Máté Stark[1], Luise Renz[1], Affan Safeer[3], Daniel Jansen[3], Thomas Michely[3], Jeison Fischer[3], Wolfgang Wernsdorfer[1,2,4], Christoph Sürgers[1], Hannu-Pekka Komsa[5], Johannes Schwenk[1,*], Wouter Jolie[3,*], Philip Willke[1,2]

[1] Physikalisches Institut (PHI), Karlsruhe Institute of Technology, Karlsruhe, Germany

[2] Center for Integrated Quantum Science and Technology (IQST), Karlsruhe Institute of Technology, Karlsruhe, Germany

[3] II. Physikalisches Institut, Universität zu Köln, Cologne, Germany

[4] Institute for Quantum Materials and Technologies (IQMT), Karlsruhe Institute of Technology, Karlsruhe, Germany

[5] Faculty of Information Technology and Electrical Engineering, University of Oulu, Oulu, Finland

[*] corresponding author: johannes.schwenk@kit.edu , wjolie@ph2.uni-koeln.de

[†] These authors contributed equally

**Abstract**

Individual spin defects in solids are promising building blocks for quantum technologies, but their deterministic creation, individual addressability, and operation near surfaces remain major challenges. Two-dimensional materials provide an attractive alternative, as their single-layer thickness enables direct atomic-scale access to defect states. Here, we demonstrate single-spin control of solid-state defects in a two-dimensional semiconductor by a combination of scanning tunneling microscopy and electron spin resonance. We create and manipulate individual sulfur vacancies and carbon substitution defects in monolayer molybdenum disulfide and characterize their spin dynamics, including coherent control, at the single-defect level. Using atomic manipulation, we further engineer and probe spin-spin interactions between defect pairs. Our results demonstrate deterministic creation, addressability, coherent manipulation, and controlled coupling of individual spin defects within a single experimental platform. This establishes atomically engineered spin defects in two-dimensional semiconductors as a versatile class of controllable solid-state quantum systems and opens a route towards tailored quantum sensing experiments.

**Introduction**

Solid-state spin defects provide localized and well-isolated quantum states for sensing, communication, and information processing[1,2]. Prominent examples include donor spins in silicon[3,4], nitrogen-vacancy centers in diamond[5,6], and rare-earth ions in crystalline hosts[7]. These quantum systems combine long coherence times with optical or electrical addressability, yet deterministic placement[8] as well as individual control[1] remain major challenges. Moreover, their performance often degrades near surfaces – a critical limitation for many sensing and device applications[9].

Two-dimensional (2D) materials offer an attractive alternative to conventional solid-state hosts: Their single-layer atomic thickness places defects directly at the surface and still preserve their quantum properties. In particular, 2D semiconductors such as transition metal dichalcogenides (TMDs) provide a broad range of electronic and optical behaviors, chemical tunability, and gate response[10]. Despite these advantages, coherent control of (individual) spin defects in 2D materials has remained largely unexplored. One notable exception is hexagonal boron nitride, where spin defects can be controlled via optically detected magnetic resonance[11-15]. Still, the microscopic nature of the spin defects is often difficult to determine[16]. For 2D semiconductors, the realization of spin defects as controllable quantum objects has so far been primarily limited to theoretical work[17-19].

Scanning tunneling microscopy (STM) provides atomic-scale access to defects in 2D materials and has revealed their electronic structure in great detail, particularly for defects in TMDs[20-24]. When combined with electron spin resonance (ESR), STM enables the detection[25,26] and coherent control[27-30] of individual spins on surfaces. This capability has established single spin resonance using STM as a powerful technique for atomically precise spin control with exceptional spatial, temporal, and energy resolution. Until now, however, this has been restricted to adatoms[25] and molecular spins[29,31] *on* surfaces, which greatly limits its potential for prototyping real-world solid-state device architectures.

In this work, we probe individual solid-state spin defects in a 2D semiconductor on the atomic scale: Using monolayer molybdenum disulfide ($MoS_2$) supported on graphene (Gr) on Ir(111), we create, characterize and coherently control individual sulfur vacancies and carbon substitution defects with atomic precision, all in one platform. We measure their spin relaxation and coherence times, demonstrate coherent Rabi and Ramsey dynamics, and engineer tunable spin-spin interactions by assembling defect dimers. Our results establish monolayer TMDs as a platform for atomically controllable solid-state spin defects and expand ESR-STM from surface-bound adatoms and molecules to genuine solid-state quantum defects.

**Electronic Structure of Spin Defects in Monolayer $MoS_2$**

Figure 1 introduces the sample system and the atomic structure of individual defect centers. Monolayer $MoS_2$ was grown on Gr/Ir(111) following established procedures[32,33] and transferred under ultra-high vacuum conditions from the growth chamber (Cologne) to the ESR-STM setup (Karlsruhe, $T = 50$ mK, Fig. 1a). In other works on 2D semiconductors, defect creation is realized via the growth process[21], ion sputtering[34] or adatom manipulation[24]. Here, we instead established protocols to create atomic defects locally, on-demand and *in-situ* using high-voltage STM tip pulses[35] (see Supplementary Fig. 1 for defect creation). This results in several distinct defect types distributed across MoS$_2$ islands (Fig. 1b). The two most common ones are shown in Fig. 1c: a negatively charged sulfur vacancy (right, $V_S^-$) and a negatively charged carbon substitution (left, $C_S^-$).

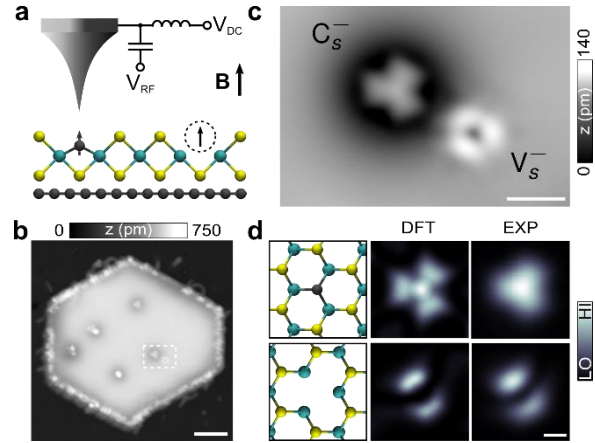

**Fig. 1. Solid-state spin defects in monolayer MoS₂. a,** Schematic illustration of MoS$_2$ on Gr/Ir(111) probed via STM (black: C; yellow: S; cyan: Mo). A carbon substitution defect ($C_S^-$, black) and a sulfur vacancy ($V_S^-$, dotted circle) are indicated with their spin. An additional RF voltage $V_{RF}$ as well as a magnetic field is employed for ESR-STM experiments. **b,** STM image of a MoS$_2$ island. Multiple atomic defects are created by STM tip manipulation (scale bar: 5 nm). **c,** Close-up STM image [marked area in b] of $V_S^-$ and $C_S^-$ [scale bar: 1 nm; Setpoints: b-c, $V_{DC} = 1$ V, $I = 20$ pA]. **d,** Left: Top view models of the defects (Top-layer only). d$I$/d$V$ orbital maps of the highest occupied defect orbital states (image size: 1.7 nm × 1.7 nm, -170 mV for $V_S^-$ and -480 mV for $C_S^-$) along with corresponding DFT-calculations.

We characterize the electronic structure of the defects using scanning tunneling spectroscopy (STS) and density functional theory (DFT) calculations (Fig. 1d and Supplementary Fig. 2 and 3). In both cases, the defect orbitals result from the $C_{3v}$ symmetry, that leads to one $A_1$ and two $E_{1,2}$ states with orbital degeneracy. In the case of $V_S^-$, spin-orbit coupling, charging, and Jahn-Teller distortion needs to be considered, which yields a highest occupied defect orbital (HODO) with a single unpaired spin in one of the $E_{1,2}$ states[20,22-24,36] while the lower lying $A_1$ states are fully occupied. This order is reversed in the case of $C_S^-$ leading to a single open-shell defect configuration with one spin in the $A_1$ orbital. The corresponding d$I$/d$V$ orbital maps of the HODOs are in excellent agreement with DFT results (Fig. 1d) and prior work[20,22-24,36,37]. Thus, while the microscopic origin of the sulfur vacancy and the carbon substitution is quite different, they both host well-defined, singly occupied defect orbitals that can serve as controllable spin $S = 1/2$ quantum states.

**Magnetic Properties and Spin Dynamics of the Defect Centers**

To probe the spin states of the defects, we perform magnetic-field-dependent d$I$/d$V$ spectroscopy (Fig. 2a). At low voltages, these measurements show an emerging step-like feature which we attribute to inelastic electron tunneling spectroscopy (IETS) excitations from the magnetic ground state to the excited state[38]. The step position $|eV_{DC}| = g\mu_B B$ yields a g-factor of ~2 for both spin centers (see Supplementary Fig. 4a for $C_S^-$). For $V_S^-$, the data agrees with the splitting of a Kondo resonance in magnetic fields observed in MoS$_2$/Au(111)[34]. Moreover, spin contrast maps (Fig. 2b; see Supplementary Fig. 4b-e for experimental details) reveal that the spatial spin density follows the shape of the HODO, leading to an extended spin signal of nearly one nanometer. According to these measurements, both defects behave as effective spin $S = 1/2$ systems with nearly-free-electron magnetic moments.

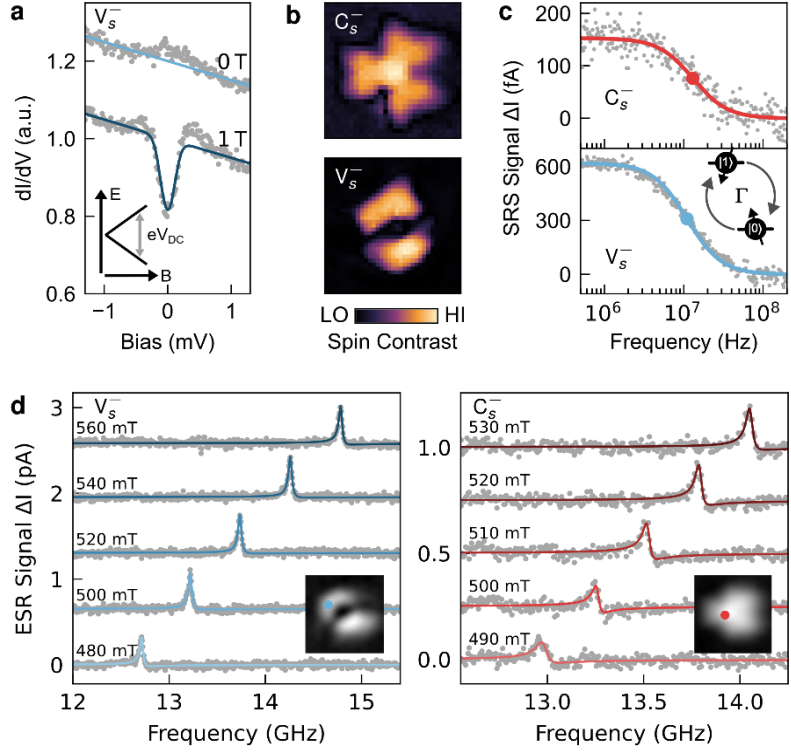

**Fig. 2. Magnetic properties and spin dynamics of individual defect centers. a**, d$I$/d$V$ spectra of $V_S^-$, that show a Zeeman splitting at finite magnetic field probed via inelastic electron tunneling excitations at $|eV_{DC}| = g\mu_B B$. **b**, Spatial distribution of the spin signal of both $C_S^-$ and $V_S^-$ revealed by spin-contrast maps $[|d^2I/dV^2|]$ acquired with a spin-polarized tip[39], 1.1 nm × 1.1 nm, see Supplementary Fig. 4b-e for details]. **c**, Stochastic resonance spectroscopy signal $\Delta I(f)$ for $C_S^-$ (red) and $V_S^-$ (blue). Fits to Eq. (1) yield spin relaxation rates between the ground and excited states [see inset, $\Gamma = (82 \pm 4)$ μs$^{-1}$ for $C_S^-$ and $(69 \pm 1)$ μs$^{-1}$ for $V_S^-$]. **d**, ESR-STM spectra $\Delta I$ of single $V_S^-$ and $C_S^-$ at different magnetic fields (taken at positions indicated in the inset). Blue and red lines are fits to asymmetric Lorentzians. [Setpoints: a, $I = 100$ pA, $V_{DC} = 5$ mV, $V_{mod} = 50$ μV; b, $V = 10$ mV & 5 mV; $I = 200$ & 500 pA; $V_{mod} = 0.8$ & 0.5 mV; $B = 0$ T; c, $I \approx 5$ pA, $|V_{DC}| = 2$ mV, $V_{AC} = 2$ mV, $B = 0$ T; d, $V_S^-$: $I = 10$ pA, $V_{DC} = 20$ mV, $V_{RF} = 5$ mV; $C_S^-$: $I = 5$ pA, $V_{DC} = 20$ mV, $V_{RF} = 10$ mV].

To quantify the spin dynamics of individual defects, we employ stochastic resonance spectroscopy (SRS)[40,41] as well as ESR-STM[25], both using magnetic STM tips (see Supplementary Note 1 for a detailed SRS analysis). In SRS, an oscillating bias voltage $V_{AC}(f)$ induces stochastic transitions between the magnetic ground and excited state, that result in a frequency-dependent change in tunneling current

$$\Delta I(f) = I_{\mathrm{SR}} \cdot \frac{\Gamma^2}{\Gamma^2+(2\pi f)^2} \quad (1)$$

where $I_{\mathrm{SR}}$ is the amplitude of the SRS response. This allows us to access nanosecond spin dynamics, since the spin transition rate $\Gamma \approx 1/T_1$ reflects the relaxation time $T_1$ (Fig. 2c)[41]. Fits to the SRS response yield relaxation times of $T_1 = (15 \pm 1)$ ns for $V_S^-$ and $T_1 = (12 \pm 1)$ ns for $C_S^-$. We find that the main limitation to $T_1$ is the tunneling current $I$[42]: $\Gamma$ increases with increasing $I$ (Supplementary Fig. 5), indicating that inelastic scattering of tunneling electrons induces spin flips. In contrast, extrapolating to zero current yields current-independent relaxation rates of $\Gamma_0^{V_S^-} = (23 \pm 1)$ MHz and $\Gamma_0^{C_S^-} = (31 \pm 9)$ MHz, corresponding to relaxation times $T_1$ of approximately 43 ns and 32 ns, respectively. These values are likely dominated by electron scattering with the conductive Gr/Ir(111) substrate[27,42] (Supplementary Note 1).

We next perform ESR-STM measurements by detecting the change $\Delta I$ in tunneling current induced by an applied radio-frequency voltage $V_{\mathrm{RF}}$ as a function of radiofrequency $f$[25]. In Fig. 2d, a pronounced resonance peak is observed at frequency $f_0$, which shifts linearly with magnetic field according to

$$hf_0 = 2\mu_{\mathrm{eff}} B. \quad (2)$$

From the field dependence, we extract effective magnetic moments $\mu_{V_S^-} = (0.992 \pm 0.011)\,\mu_B$ and $\mu_{C_S^-} = (0.975 \pm 0.007)\,\mu_B$, close to the free-electron value in both cases (see Supplementary Fig. 6 for effective magnetic moments evaluation). We attribute the deviation from the latter to the residual influence of spin-orbit coupling and Kondo coupling to the substrate electron bath[43].

**Coherent control of a single sulfur vacancy spin**

Having established the spin properties of the defect centers, we next probe the coherent dynamics of a single $V_S^-$ spin using pulsed ESR-STM[27]: Figure 3a shows measurements of the change in tunneling current $\Delta I$ as a function of RF pulse width $\tau$, that demonstrate coherent Rabi oscillations of the spin defect. The Rabi frequency $\Omega$ increases linearly with $V_{\mathrm{RF}}$ (inset; see Methods for analysis details), consistent with single-spin control of adatoms[27] and molecules[29]. From this data, we extract a decoherence time $T_2^{\mathrm{Rabi}} = (22 \pm 1)$ ns ($V_{\mathrm{RF}} = 40$ mV) and a minimal $\pi$-pulse duration $T_\pi^{\min} \approx 14$ ns ($V_{\mathrm{RF}} = 50$ mV). Detuning measurements further confirm coherent spin driving: mapping the Rabi sequence as a function of detuning $\delta = f - f_0$ (Fig. 3b) yields the characteristic chevron pattern and the extracted Rabi rates follow the expected relation $\Omega = \sqrt{\Omega_0^2 + \delta^2}$. To quantify spin relaxation and decoherence, we performed relaxometry and Ramsey experiments (Fig. 3c and d). The relaxation measurement using a two-$\pi$-pulse sequence[30] gives $T_1 = (27 \pm 3)$ ns. Ramsey fringes obtained from $\pi/2$–$\pi/2$ pulse sequences show oscillatory decay with $T_2^* = (13.5 \pm 0.6)$ ns (see Methods for analysis details). Since the lifetime remains close to those obtained by SRS (Fig. 2c), we conclude that the main limitation to $T_1$ is also here inelastic scattering with tunneling electrons. Concerning $T_2^*$, additionally dephasing is potentially introduced by the variations in the tip magnetic field[27].

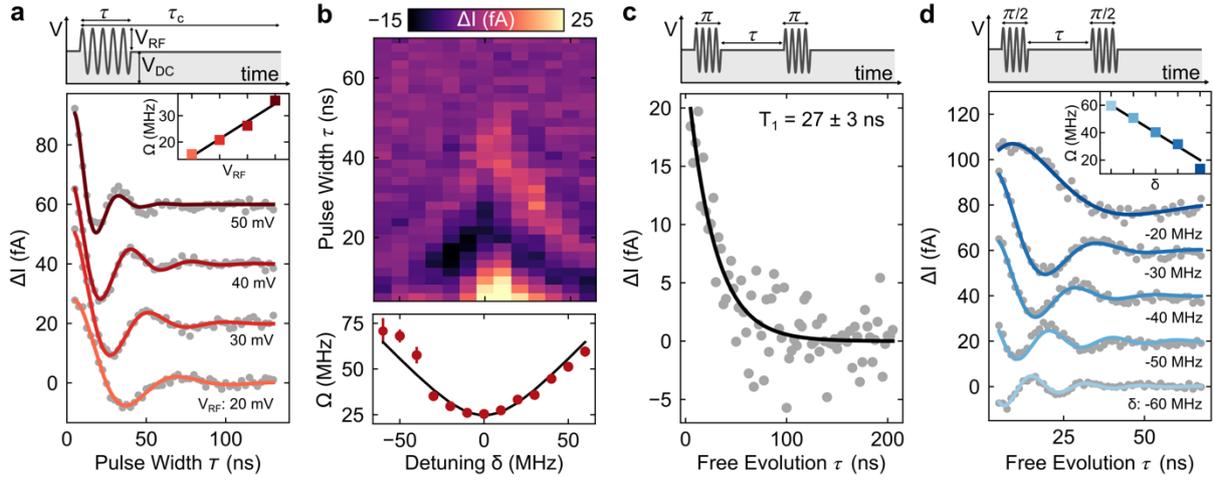

**Fig. 3. Coherent control of a single sulfur vacancy spin. a**, Rabi oscillation measurements on a single $V_S^-$. Top: Pulse scheme illustrating the DC bias $V_{DC}$, RF amplitude $V_{RF}$, RF pulse width $\tau$ and cycle time $\tau_c$. Bottom: Change in tunneling current $\Delta I(\tau)$ for different $V_{RF}$. Solid lines are fits to the data (See Methods). Inset: extracted Rabi rates $\Omega$ show a linear dependence on $V_{RF}$. **b**, Rabi oscillations measured as a function of detuning $\delta = f - f_0$. Top: $\Delta I(\tau, \delta)$ displaying the characteristic chevron pattern of coherent spin driving. Bottom: Extracted Rabi rates following the expected detuning dependence $\Omega(\delta) = \sqrt{\Omega_0^2 + \delta^2}$ (solid line). **c**, Spin relaxometry measurement. Top: Pulse scheme using two $\pi$-pulses separated by a free evolution time $\tau$. Bottom: $\Delta I(\tau)$ with an exponential fit yielding a spin relaxation time $T_1 = (27 \pm 3)$ ns. **d**, Ramsey interference measurement. Top: Pulse scheme. Bottom: $\Delta I(\tau)$ for different detuning $\delta$, showing oscillatory decay of the Ramsey fringes. Fits yield an average dephasing time $T_2^* = (13.5 \pm 0.6)$ ns, (See Methods). Inset: Extracted oscillation frequencies $\Omega$ as a function of $\delta$ (colors indicating the detuning) matching the expected linear dependence (black line) [Setpoints: All measurements were performed in open loop with $B = 527$ mT, $I = 4.5$ pA, $V_{DC} = 20$ mV and $f_0 = 14.1$ GHz; $V_{RF} = 40$ mV for b-d; a-b, $\tau_c = 250$ ns; c, $\tau_c = 600$ ns, $\tau_\pi = 22$ ns; d, $\tau_c = 150$ ns, $\tau_{\pi/2} = 11$ ns].

**Engineering spin-spin interactions in defect dimers**

Lastly, we utilize the atomic precision of STM to engineer defect dimers via tip-assisted manipulation (Supplementary Fig. 1) and to investigate their magnetic coupling: We assemble pairs of $C_S^-$ on well-defined lattice sites and probe their mutual spin-spin interaction using ESR-STM and IETS (Fig. 4). For sufficiently separated defects in a (2,2) configuration (given in multiples of the MoS$_2$ lattice vectors), the system behaves as two weakly interacting $S = 1/2$ spins (Fig. 4a). ESR-STM reveals two distinct resonance peaks, consistent with additional Heisenberg exchange interaction described by $H_{ex} = J\, S_1 \cdot S_2$. In the present case, the (ferromagnetic) exchange $J \approx -400$ MHz (~$-2$ μeV) is small compared to the Zeeman energy, a typical regime for coupled two-qubit systems. This leads to a frequency splitting of $\Delta f = |J|/h$, which can be interpreted as an effective field generated by the remote defect switching between its two possible spin orientations (Fig. 4a inset, Supplementary Fig. 7 and

Supplementary Note 3). When the defects are positioned in a (1,3) configuration (Fig. 4b), the exchange interaction increases significantly and reaches $J \approx +354$ µeV. At this magnitude, the exchange becomes directly visible in d$I$/d$V$ spectra as IETS excitations (Fig. 4b inset), which unlike in Fig. 2a persist even in the absence of magnetic field $B$. Notably, the increase in $J$ occurs despite a larger absolute geometric separation between the $C_S^-$ defects: Since the orbital overlap is increased in this angle, the magnetic exchange is increased as well.

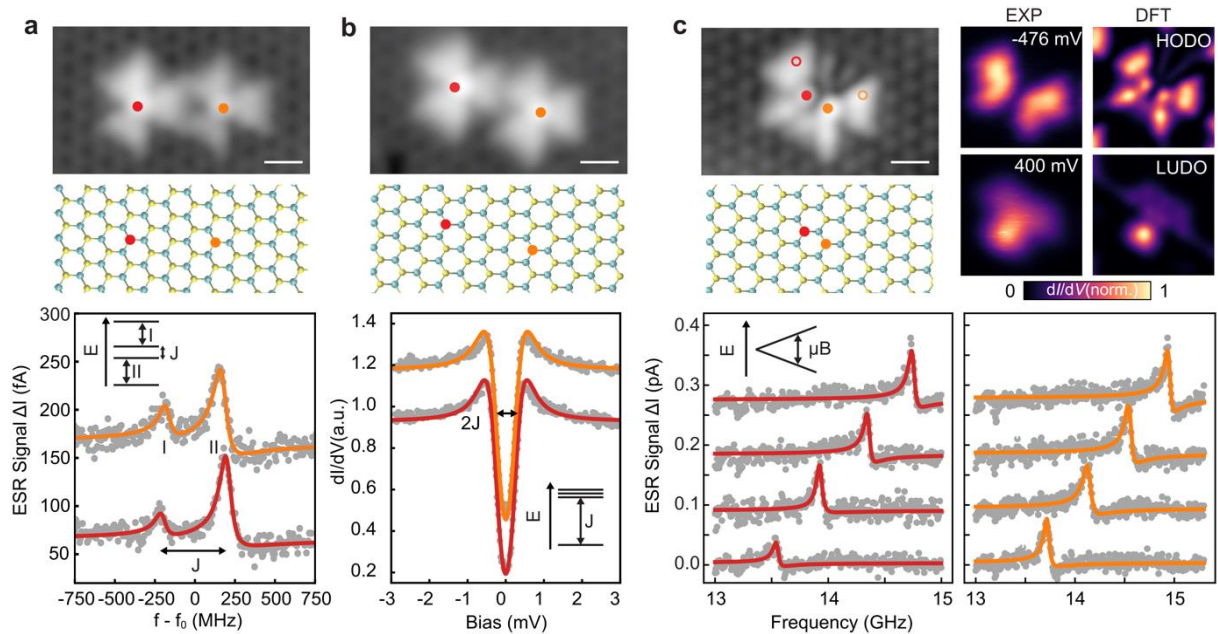

**Fig. 4. Engineering spin-spin interaction in carbon substitution dimers. a – c**, Atomic-scale assembly and magnetic characterization of $C_S^-$ with different separations. Top: STM topographies (scale bars: 0.5 nm); middle: atomic models; bottom: spectroscopic signatures. **a**, For a weakly coupled dimer in a (2,2) configuration, ESR-STM reveals two distinct resonances corresponding to individual spins, split by a small (ferromagnetic) exchange interaction $J$ of $\sim -400$ MHz (inset, see also Supplementary Fig. 7). **b**, For a (1,3) configuration, the interaction increases drastically to $J = 354$ µeV, which becomes directly visible as IETS in low-bias d$I$/d$V$. **c**, At the closest possible (0,1) configuration, strong orbital overlap and on-site Coulomb repulsion led to the formation of a new electronic state, $(2C_S)^-$, hosting a single unpaired electron. Constant-height d$I$/d$V$ maps and DFT calculations reveal modified orbitals (upper right), while ESR-STM (bottom left and right) detects the response of a single isolated spin $S = 1/2$ without exchange splitting (Spectra are taken at the open circle positions to emphasize the delocalized nature of the defect spin). Thus, $(2C_S)^-$ mimics the formation of a hydrogen molecule ion and gives rise to emergent states beyond a coupled-spin model. [Setpoints: a, $I = 30$ pA, $V_{DC} = 30$ mV, $V_{RF} = 10$ mV, $B = 520$ mT, $f_0 = 14.264$ / $14.222$ GHz; b, $I = 200$ pA, $V_{DC} = 10$ mV, $V_{mod} = 40$ µV, $B = 0$ T; c, Bottom left: $I = 10$ pA, $V_{DC} = 30$ mV, $V_{RF} = 12$ mV, Bottom right: $I = 15$ pA, $V_{DC} = 30$ mV, $V_{RF} = 10$ mV. Difference of $\Delta f \sim 200$ MHz between the two measurements is attributed to the residual tip magnetic field[44]].

A qualitatively different regime is reached for the closest possible configuration, (0,1), shown in Fig. 4c. Here, magnetic-field-dependent ESR-STM spectra acquired at different positions

(Fig. 4c, bottom) exhibit the response of a single isolated $S = 1/2$ spin [ $\mu_{(2C_S)^-}$ = $(0.975 \pm 0.004)$ $\mu_B$], with no observable exchange splitting. Consistently, no IETS signal is detected in d$I$/d$V$ in zero magnetic field. To shed light on the origin of these results, we performed d$I$/d$V$ maps of the HODO and LUDO (Fig. 4c, upper right; see Supplementary Fig. 8a,b for d$I$/d$V$ spectrum and additional maps). These maps reveal orbital shapes that differ markedly from those of isolated $C_S^-$ defects (Fig. 1d). DFT calculations (Supplementary Fig. 8c-e and Supplementary Note 2) show that this is a combined effect of strong orbital overlap and large on-site Coulomb repulsion, which lead to a change in the orbital structure. This also suppresses double occupancy, which leaves only a single electron localized on the dimer, rationalizing the ESR measurements. As a result, the new defect $(2C_S)^-$ undergoes a qualitative transition characterized by orbital states that are distinct from those of its two $C_S^-$ constituents. In close analogy to the formation of the hydrogen molecule ion $H_2^+$ from two hydrogen atoms, $(2C_S)^-$ constitutes an electronically and magnetically distinct entity rather than a strongly coupled spin pair.

**Discussion**

A central challenge for future progress is to further enhance spin coherence. While coherent manipulation is already achieved, prolonging spin lifetimes will require to reduce inelastic electron scattering, that stems from the presence of the tunneling current and proximity to the conducting electrodes. On the detection side, this can be addressed by remote spin readout schemes[28] or by force-based detection[45] that decouple spin readout from electronic transport. On the materials side, substrate-induced relaxation can be mitigated by introducing additional non-conductive decoupling layers (e.g. bilayer $MoS_2$ or hexagonal boron nitride), by employing insulating substrates such as silicon carbide[22,23] or by using freestanding $MoS_2$[46].

Equally important is the deterministic creation and placement of defects using STM-based atom manipulation. This capability opens a route toward atomically engineered spin structures in two dimensions, analogous to phosphorus donor-based architectures in silicon[8], but with the additional advantage of in-situ tuning, characterization, and control at the single-spin level. In this context, extending ESR-STM from adatoms and molecular spins to solid-state spin defects significantly broadens its scope and establishes it as a general tool for coherent control in complex solid-state systems.

More generally, the combination of atomic-scale addressability, engineered spin-spin interactions, and surface accessibility makes spin defects in 2D materials particularly well-suited for integration into functional nanoscale devices. These attributes provide a foundation for exploring nanoscale quantum sensing, engineered spin networks, and model quantum systems. Together, our results position spin defects in 2D materials as a viable platform for solid-state quantum nanotechnologies.

**Methods**

**Sample preparation**
Monolayer $MoS_2$ on graphene/Ir(111) is prepared by molecular beam epitaxy in an ultrahigh-vacuum (UHV) chamber (base pressure = $1 \times 10^{-10}$ mbar). First, Ir(111) is cleaned by cycles of

1 keV Ar$^+$ ion sputtering and flash annealing to 1510 K. To obtain single layer graphene on Ir(111), clean Ir(111) is exposed to 1 × 10$^{-7}$ mbar ethylene at room temperature for 120 s. Afterwards, the sample is shortly annealed to 1470 K without ethylene. Subsequently, the sample is exposed to 3 × 10$^{-7}$ mbar ethylene at 1370 K for 600 s to obtain a full graphene monolayer on Ir(111)[47]. The sample is then exposed to a Mo flux from an e-beam evaporator simultaneously with a S flux obtained by decomposing pyrite (FeS$_2$) in a Knudsen cell (distance ~8 cm). The S flux is characterized by a distant ion gauge measuring a S pressure of 1 × 10$^{-8}$ mbar. The sample is subsequently annealed at 1000 K for 300 s in the same sulfur flux, following Ref. [48].

For ESR-STM measurements, the sample is transferred from the preparation chamber to the ESR-STM chamber using an ultrahigh vacuum suitcase (base pressure = 5 × 10$^{-10}$ mbar). All measurements were done in a Unisoku USM1600 STM inside a homebuilt dilution refrigerator with a base temperature of 50 mK.

**Spin polarized tips**

Fe atom deposition is carried out by electron-beam evaporation (3 s, ion current ~5 nA at ~30 cm distance) onto the cooled (<100 K) MoS$_2$/(Gr)/Ir(111) sample. All ESR and SRS experiments were carried out using spin-polarized tips. These were prepared by transferring individual Fe adatoms adsorbed on either graphene or MoS$_2$ onto the PtIr tip by STM vertical atom manipulation. Spin polarization was then verified through the asymmetry in d$I$/d$V$ measurements around zero bias on $V_S^-$ or $C_S^-$ defects (Supplementary Fig. 4b-e). Note that while the bias voltage was applied to the STM tip, all bias signs were inverted in the manuscript to follow the conventional definition of bias voltage with respect to the sample bias.

**ESR and pulsed ESR**

The RF voltage was added to the tip side of the junction via a cryogenic diplexer (Marki Microwave MDPX-0305) located at the 1 K stage of the cryostat using an RF generator (Anritsu MG3694C). For pulsed ESR measurements, we followed the pulse schemes introduced in Ref. [27]. The continuous wave RF signal from the RF generator was used as the local oscillator of an IQ mixer (Marki MMIQ0626H). Time-dependent $I(t)$ and $Q(t)$ from an arbitrary waveform generator (Zurich Instruments HDAWG) drove the I and Q ports to gate the RF signal and apply phase-coherent pulse sequences ($\pi$ or $\pi/2$ pulses). A 3-dB attenuator was added (at room temperature) before the cryogenic diplexer. The ESR signal in the tunnelling current was read out with a digital lock-in (Stanford Research Systems SR860) using on/off modulation at 323 Hz. The RF pulses were only present in the Lockin A-cycle. Therefore, the measured signal presents an average of the spin state-related tunnel current. The DC background is applied over the whole cycle time $\tau_c$. we applied pulse sequences with on-time $\tau$ and off-time $\tau_{off}$ keeping the sum ($\tau_c = \tau + \tau_{off}$) fixed.

Rabi oscillations were analyzed by using a damped cosine model (solid line in Fig. 3a) with an exponential envelope, $\Delta I(\tau) = A \cdot \cos(2\pi\Omega\tau+\phi) \exp[-\tau/T_2^{Rabi}]$. Here, $\tau$ is the pulse width, $A$ is the maximum change in current and $T_2^{Rabi}$ is the Rabi decoherence time. A linear background caused by current rectification with increasing $\tau$ was subtracted from the data [27]. Rabi frequency $\Omega$ extracted from the fits as a function of $V_{RF}$. $T_2^{Rabi}$ was then extracted from the fits in Fig. 3a.

Ramsey interference measurements were analyzed by using the standard Ramsey model (solid line in Fig. 3d) $\Delta I(\tau) = A \cdot \cos(2\pi\Omega\tau + \phi) \exp(-\tau/T_2^*) + I_0$, where $A$ is the amplitude, $\phi$ a phase offset, and $T_2^*$ the inhomogeneous dephasing time. Oscillation frequencies $\Omega$ extracted from the fits as a function of detuning $\delta$. $T_2^*$ was then extracted as a function of detuning.

**SRS measurements**

For SRS measurements, the low-frequency offset voltage was combined with the high-frequency modulation using an external bias-tee (Mini-circuits ZFBT-6GW+) and it was applied to the tip-side of the tunnel junction via the low frequency input of the cryogenic diplexer. The AC modulation signal was produced by an arbitrary waveform generator (Zurich Instruments HDAWG). The SRS signal was read out with a digital lock-in (Stanford Research Systems SR860) using on/off modulation at 323 Hz.

**Computational methods**

Density functional theory (DFT) calculations are all performed within the projector-augmented wave formalism implemented in the VASP software package [49-51]. For the presented results in the figures, Heyd-Scuseria-Ernzerhof (HSE06) hybrid functionals and a mixing parameter of 10% were used [52,53]. This was found to yield energy differences between occupied and unoccupied defect states, in closer agreement with experiments, as compared to e.g. those from the Perdew-Burke-Ernzerhof (PBE) functional [54] or HSE06 with standard mixing parameter of 25%. Hybrid functionals also improve on the predicted band gap value, which is found to be 1.91 eV for HSE-10%. A plane-wave basis with a 500 eV cutoff was employed. Both defects were modeled within a 6×6 supercell and the Brillouin zone was sampled using a 2×2 k-point-mesh. The structural relaxations were carried out in spin-polarized form, followed by single-shot non-collinear calculations with spin-orbit coupling. STM images were simulated using Tersoff-Hamann approach [55], essentially corresponding to evaluation of a local DOS. In practice, the squared wave functions of defect states were extracted using the VASPKIT package [56], and from that we took a plane at a constant height from the surface.

## References

1  Wolfowicz, G. *et al.* Quantum guidelines for solid-state spin defects. *Nature Reviews Materials* **6**, 906-925 (2021). https://doi.org/10.1038/s41578-021-00306-y
2  Awschalom, D. D., Bassett, L. C., Dzurak, A. S., Hu, E. L. & Petta, J. R. Quantum Spintronics: Engineering and Manipulating Atom-Like Spins in Semiconductors. *Science* **339**, 1174-1179 (2013). https://doi.org/doi:10.1126/science.1231364
3  Morello, A. *et al.* Single-shot readout of an electron spin in silicon. *Nature* **467**, 687-691 (2010). https://doi.org/10.1038/nature09392
4  Watson, T. F. *et al.* Atomically engineered electron spin lifetimes of 30 s in silicon. *Science Advances* **3**, e1602811 (2017). https://doi.org/doi:10.1126/sciadv.1602811

5 Balasubramanian, G. *et al.* Nanoscale imaging magnetometry with diamond spins under ambient conditions. *Nature* **455**, 648-651 (2008). https://doi.org/10.1038/nature07278
6 Thiel, L. *et al.* Probing magnetism in 2D materials at the nanoscale with single-spin microscopy. *Science* **364**, 973-976 (2019). https://doi.org/doi:10.1126/science.aav6926
7 Yin, C. *et al.* Optical addressing of an individual erbium ion in silicon. *Nature* **497**, 91-94 (2013). https://doi.org/10.1038/nature12081
8 Kiczynski, M. *et al.* Engineering topological states in atom-based semiconductor quantum dots. *Nature* **606**, 694-699 (2022). https://doi.org/10.1038/s41586-022-04706-0
9 Marchiori, E. *et al.* Nanoscale magnetic field imaging for 2D materials. *Nature Reviews Physics* **4**, 49-60 (2022). https://doi.org/10.1038/s42254-021-00380-9
10 Liu, X. & Hersam, M. C. 2D materials for quantum information science. *Nature Reviews Materials* **4**, 669-684 (2019). https://doi.org/10.1038/s41578-019-0136-x
11 Gottscholl, A. *et al.* Initialization and read-out of intrinsic spin defects in a van der Waals crystal at room temperature. *Nature Materials* **19**, 540-545 (2020). https://doi.org/10.1038/s41563-020-0619-6
12 Chejanovsky, N. *et al.* Single-spin resonance in a van der Waals embedded paramagnetic defect. *Nature Materials* **20**, 1079-1084 (2021). https://doi.org/10.1038/s41563-021-00979-4
13 Stern, H. L. *et al.* A quantum coherent spin in hexagonal boron nitride at ambient conditions. *Nature Materials* **23**, 1379-1385 (2024). https://doi.org/10.1038/s41563-024-01887-z
14 Gao, X. *et al.* Single nuclear spin detection and control in a van der Waals material. *Nature* **643**, 943-949 (2025). https://doi.org/10.1038/s41586-025-09258-7
15 Whitefield, B. *et al.* Narrowband quantum emitters in hexagonal boron nitride with optically addressable spins. *Nature Materials* (2026). https://doi.org/10.1038/s41563-025-02458-6
16 Mendelson, N. *et al.* Identifying carbon as the source of visible single-photon emission from hexagonal boron nitride. *Nature Materials* **20**, 321-328 (2021). https://doi.org/10.1038/s41563-020-00850-y
17 Li, S., Thiering, G., Udvarhelyi, P., Ivády, V. & Gali, A. Carbon defect qubit in two-dimensional $WS_2$. *Nature Communications* **13**, 1210 (2022). https://doi.org/10.1038/s41467-022-28876-7
18 Tsai, J.-Y., Pan, J., Lin, H., Bansil, A. & Yan, Q. Antisite defect qubits in monolayer transition metal dichalcogenides. *Nature Communications* **13**, 492 (2022). https://doi.org/10.1038/s41467-022-28133-x
19 Lee, Y. *et al.* Spin-defect qubits in two-dimensional transition metal dichalcogenides operating at telecom wavelengths. *Nature Communications* **13**, 7501 (2022). https://doi.org/10.1038/s41467-022-35048-0
20 Aliyar, T. *et al.* Symmetry Breaking and Spin–Orbit Coupling for Individual Vacancy-Induced In-Gap States in $MoS_2$ Monolayers. *Nano Letters* **24**, 2142-2148 (2024). https://doi.org/10.1021/acs.nanolett.3c03681
21 Schuler, B. *et al.* How substitutional point defects in two-dimensional $WS_2$ induce charge localization, spin–orbit splitting, and strain. *ACS Nano* **13**, 10520-10534 (2019). https://doi.org/10.1021/acsnano.9b04611
22 Schuler, B. *et al.* Large Spin-Orbit Splitting of Deep In-Gap Defect States of Engineered Sulfur Vacancies in Monolayer $WS_2$ *Physical Review Letters* **123**, 076801 (2019). https://doi.org/10.1103/PhysRevLett.123.076801

23  Schuler, B. *et al.* Electrically driven photon emission from individual atomic defects in monolayer $WS_2$. *Science Advances* **6**, eabb5988 (2020). https://doi.org/doi:10.1126/sciadv.abb5988
24  Jansen, D. *et al.* Tip-induced creation and Jahn-Teller distortions of sulfur vacancies in single-layer $MoS_2$. *Physical Review B* **109**, 195430 (2024). https://doi.org/10.1103/PhysRevB.109.195430
25  Baumann, S. *et al.* Electron paramagnetic resonance of individual atoms on a surface. *Science* **350**, 417-420 (2015). https://doi.org/doi:10.1126/science.aac8703
26  Willke, P. *et al.* Probing quantum coherence in single-atom electron spin resonance. *Science Advances* **4**, eaaq1543 (2018). https://doi.org/doi:10.1126/sciadv.aaq1543
27  Yang, K. *et al.* Coherent spin manipulation of individual atoms on a surface. *Science* **366**, 509-512 (2019). https://doi.org/doi:10.1126/science.aay6779
28  Wang, Y. *et al.* An atomic-scale multi-qubit platform. *Science* **382**, 87-92 (2023). https://doi.org/doi:10.1126/science.ade5050
29  Willke, P. *et al.* Coherent Spin Control of Single Molecules on a Surface. *ACS Nano* **15**, 17959-17965 (2021). https://doi.org/10.1021/acsnano.1c06394
30  Kovarik, S. *et al.* Spin torque–driven electron paramagnetic resonance of a single spin in a pentacene molecule. *Science* **384**, 1368-1373 (2024). https://doi.org/10.1126/science.adh4753
31  Zhang, X. *et al.* Electron spin resonance of single iron phthalocyanine molecules and role of their non-localized spins in magnetic interactions. *Nature Chemistry* **14**, 59-65 (2022). https://doi.org/10.1038/s41557-021-00827-7
32  N'Diaye, A. T., Coraux, J., Plasa, T. N., Busse, C. & Michely, T. Structure of epitaxial graphene on Ir(111). *New Journal of Physics* **10**, 043033 (2008). https://doi.org/10.1088/1367-2630/10/4/043033
33  Hall, J. *et al.* Molecular beam epitaxy of quasi-freestanding transition metal disulphide monolayers on van der Waals substrates: a growth study. *2D Materials* **5**, 025005 (2018). https://doi.org/10.1088/2053-1583/aaa1c5
34  Trishin, S., Lotze, C., Krane, N. & Franke, K. J. Electronic and magnetic properties of single chalcogen vacancies in $MoS_2$/Au(111). *Physical Review B* **108**, 165414 (2023). https://doi.org/10.1103/PhysRevB.108.165414
35  Hosoki, S., Hosaka, S. & Hasegawa, T. Surface modification of $MoS_2$ using an STM. *Applied Surface Science*, 643-647 (1992). https://doi.org/https://doi.org/10.1016/0169-4332(92)90489-K
36  Xiang, F. *et al.* Charge state-dependent symmetry breaking of atomic defects in transition metal dichalcogenides. *Nature Communications* **15**, 2738 (2024). https://doi.org/10.1038/s41467-024-47039-4
37  Cochrane, K. A. *et al.* Spin-dependent vibronic response of a carbon radical ion in two-dimensional $WS_2$. *Nature Communications* **12**, 7287 (2021). https://doi.org/10.1038/s41467-021-27585-x
38  Ternes, M. Spin excitations and correlations in scanning tunneling spectroscopy. *New Journal of Physics* **17**, 063016 (2015). https://doi.org/10.1088/1367-2630/17/6/063016
39  Huang, W. *et al.* Quantum spin-engineering in on-surface molecular ferrimagnets. *Nature Communications* **16**, 5208 (2025). https://doi.org/10.1038/s41467-025-60409-w
40  Hänze, M. *et al.* Quantum stochastic resonance of individual Fe atoms. *Science Advances* **7**, eabg2616 (2021). https://doi.org/doi:10.1126/sciadv.abg2616
41  Betz, N. *et al.* Stochastic Resonance Spectroscopy: Characterizing Fast Dynamics with Slow Measurements. *arXiv preprint arXiv:2412.12647* (2024).

42 Paul, W. *et al.* Control of the millisecond spin lifetime of an electrically probed atom. *Nature Physics* **13**, 403-407 (2017). https://doi.org/10.1038/nphys3965
43 Bagchi, M. *et al.* Spin-Polarized Scanning Tunneling Microscopy Measurements of an Anderson Impurity. *Physical Review Letters* **133**, 246701 (2024). https://doi.org/10.1103/PhysRevLett.133.246701
44 Willke, P., Yang, K., Bae, Y., Heinrich, A. J. & Lutz, C. P. Magnetic resonance imaging of single atoms on a surface. *Nature Physics* **15**, 1005-1010 (2019). https://doi.org/10.1038/s41567-019-0573-x
45 Sellies, L. *et al.* Single-molecule electron spin resonance by means of atomic force microscopy. *Nature* **624**, 64-68 (2023). https://doi.org/10.1038/s41586-023-06754-6
46 Krane, N., Lotze, C., Läger, J. M., Reecht, G. & Franke, K. J. Electronic Structure and Luminescence of Quasi-Freestanding $MoS_2$ Nanopatches on Au(111). *Nano Letters* **16**, 5163-5168 (2016). https://doi.org/10.1021/acs.nanolett.6b02101
47 Coraux, J. *et al.* Growth of graphene on Ir(111). *New Journal of Physics* **11**, 023006 (2009). https://doi.org/10.1088/1367-2630/11/2/023006
48 Ehlen, N. *et al.* Narrow photoluminescence and Raman peaks of epitaxial $MoS_2$ on graphene/Ir(1 1 1). *2D Materials* **6**, 011006 (2019). https://doi.org/10.1088/2053-1583/aaebd3
49 Kresse, G. & Hafner, J. Ab initio molecular dynamics for liquid metals. *Physical Review B* **47**, 558-561 (1993). https://doi.org/10.1103/PhysRevB.47.558
50 Kresse, G. & Furthmüller, J. Efficiency of ab-initio total energy calculations for metals and semiconductors using a plane-wave basis set. *Computational Materials Science* **6**, 15-50 (1996). https://doi.org/https://doi.org/10.1016/0927-0256(96)00008-0
51 Kresse, G. & Furthmüller, J. Efficient iterative schemes for ab initio total-energy calculations using a plane-wave basis set. *Physical Review B* **54**, 11169-11186 (1996). https://doi.org/10.1103/PhysRevB.54.11169
52 Heyd, J., Scuseria, G. E. & Ernzerhof, M. Hybrid functionals based on a screened Coulomb potential. *The Journal of Chemical Physics* **118**, 8207-8215 (2003). https://doi.org/10.1063/1.1564060
53 Krukau, A. V., Vydrov, O. A., Izmaylov, A. F. & Scuseria, G. E. Influence of the exchange screening parameter on the performance of screened hybrid functionals. *The Journal of Chemical Physics* **125** (2006). https://doi.org/10.1063/1.2404663
54 Perdew, J. P., Burke, K. & Ernzerhof, M. Generalized Gradient Approximation Made Simple. *Physical Review Letters* **77**, 3865-3868 (1996). https://doi.org/10.1103/PhysRevLett.77.3865
55 Tersoff, J. & Hamann, D. R. Theory of the scanning tunneling microscope. *Physical Review B* **31**, 805-813 (1985). https://doi.org/10.1103/PhysRevB.31.805
56 Wang, V., Xu, N., Liu, J.-C., Tang, G. & Geng, W.-T. VASPKIT: A user-friendly interface facilitating high-throughput computing and analysis using VASP code. *Computer Physics Communications* **267**, 108033 (2021). https://doi.org/https://doi.org/10.1016/j.cpc.2021.108033
57 Seifert, T. S., Kovarik, S., Gambardella, P. & Stepanow, S. Accurate measurement of atomic magnetic moments by minimizing the tip magnetic field in STM-based electron paramagnetic resonance. *Physical Review Research* **3**, 043185 (2021). https://doi.org/10.1103/PhysRevResearch.3.043185
58 Komsa, H.-P. & Krasheninnikov, A. V. Native defects in bulk and monolayer ${\mathrm{MoS}}_{2}$ from first principles. *Physical Review B* **91**, 125304 (2015). https://doi.org/10.1103/PhysRevB.91.125304

| | |
|---|---|
| 59 | Yang, K. *et al.* Engineering the Eigenstates of Coupled Spin-$1/2$ Atoms on a Surface. *Physical Review Letters* **119**, 227206 (2017). https://doi.org/10.1103/PhysRevLett.119.227206 |
| 60 | Veldman, L. M. *et al.* Free coherent evolution of a coupled atomic spin system initialized by electron scattering. *Science* **372**, 964-968 (2021). https://doi.org/10.1126/science.abg8223 |
| 61 | Bae, Y. *et al.* Enhanced quantum coherence in exchange coupled spins via singlet-triplet transitions. *Science Advances* **4**, eaau4159 https://doi.org/10.1126/sciadv.aau4159 |

## Acknowledgments

P.W. acknowledges funding from the Emmy Noether Programme of the Deutsche Forschungsgemeinschaft (DFG, WI5486/1-2). P.G. and P.W. acknowledge financial support from the Hector Fellow Academy (Grant No. 700001123). P.W. and K.H.A.Y. acknowledge support from the Center for Integrated Quantum Science and Technology (IQST). P.W., J.S., J.A., L.H. and L.R. acknowledge funding from the ERC Starting Grant ATOMQUANT. W.J. acknowledges funding from the DFG through priority program SPP2244 (Project No. 535290457). W.J. and T.M. acknowledge funding from DFG through CRC1238 (project number 277146847, subprojects A01 and B06).

## Contributions

P.W., W.J and J.S. conceived the experiment. K.H.A.Y, W.H., J.M., P.G., J.A., L.H., M.S., L.R., J.S., C.S., W.W. and P.W. set up the experiment and conducted the ESR-STM measurements. W.J., A.S., D.J., J.F., T.M. prepared the sample, K.H.A.Y, W.H., J.M., P.G., J.A., L.H., M.S., L.R., J.S., W.J., A.S., D.J., J.F analyzed the experimental data. H.P.K. performed the DFT calculations. P.W., K.H.A.Y, W.H., J.M., P.G., J.A., J.S., D.J., H.P.K. and W.J. wrote the manuscript with input from all authors. P.W., W.J. and J.S. supervised the project.

## Competing interests

The authors declare no competing interests.

# Supplementary Information

# Atomic-Scale Quantum Control of Individual Spin Defects in a Two-Dimensional Semiconductor

Kwan Ho Au-Yeung[1,2,†], Wantong Huang[1,†], Johanna Matusche[1,†], Paul Greule[1], Jonas Arnold[1], Lovis Hardeweg[1], Máté Stark[1], Luise Renz[1], Affan Safeer[3], Daniel Jansen[3], Thomas Michely[3], Jeison Fischer[3], Wolfgang Wernsdorfer[1,2,4], Christoph Sürgers[1], Hannu-Pekka Komsa[5], Johannes Schwenk[1*], Wouter Jolie[3*], Philip Willke[1,2]

[1]     Physikalisches Institut (PHI), Karlsruhe Institute of Technology, Karlsruhe, Germany

[2]     Center for Integrated Quantum Science and Technology (IQST), Karlsruhe Institute of Technology, Karlsruhe, Germany

[3]     II. Physikalisches Institut, Universität zu Köln, Cologne, Germany

[4]     Institute for Quantum Materials and Technologies (IQMT), Karlsruhe Institute of Technology, Karlsruhe, Germany

[5]     Faculty of Information Technology and Electrical Engineering, University of Oulu, Oulu, Finland

[*] corresponding author: johannes.schwenk@kit.edu , wjolie@ph2.uni-koeln.de

[†] These authors contributed equally

## Table of Contents

**Supplementary Figs. 1-10**

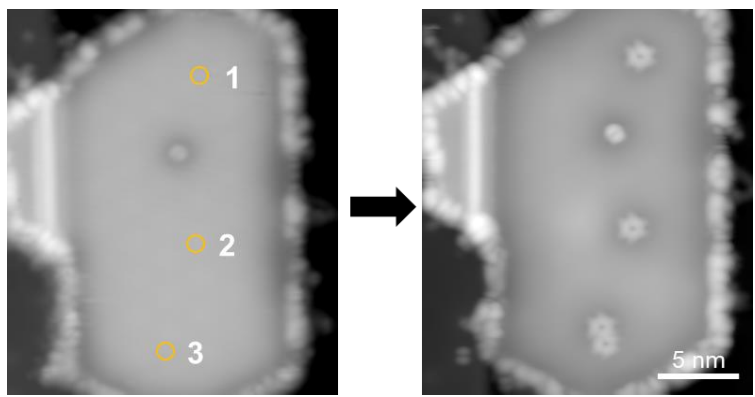

**Supplementary Fig. 1. STM tip-assisted creation of point defects on intact monolayer MoS₂ islands.** STM images acquired before and after tip-induced defect creation using voltage-pulse protocols. Defects are generated either by applying high-voltage pulses (>5 V, 100 ms, feedback off) at fixed tip position or by approaching the tip toward the surface first ($V_{DC}$ = 40 mV, $I$ = 20 pA, $\Delta z$ = 650 pm) followed by a voltage pulse (typically 1.7 V, 100 ms). Both methods enable creation of sulfur vacancies ($V_S^-$), while carbon substitutions ($C_S^-$) are preferably formed with the first method. Movement of $C_S^-$ (Fig. 4) has been performed via the first method. The latter required the incorporation of carbon, likely from nearby organic adsorbates or the underlying graphene substrate. The example shown above illustrates the creation of four $V_S^-$ defects at the three tip pulsing positions with the first method; an additional defect present prior to the sequence is also visible (Imaging parameters: $V_{DC}$ = 1 V, $I$ = 20 pA).

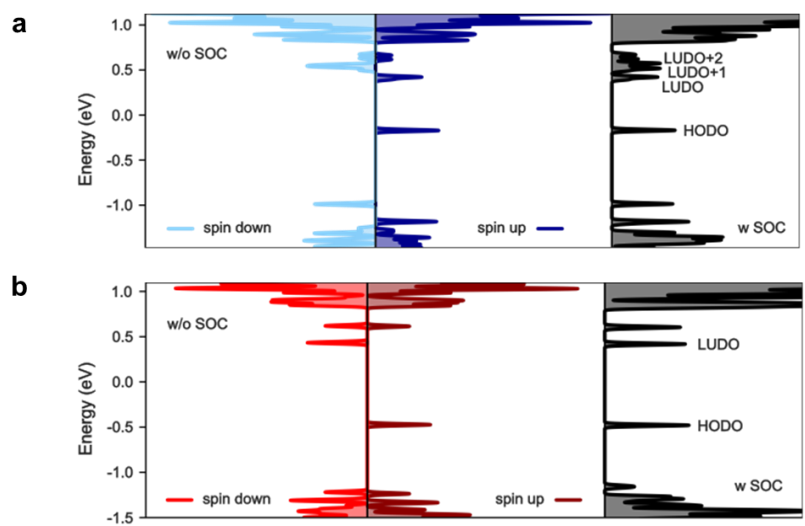

**Supplementary Fig. 2. DFT-calculated density of states** of **a,** the negatively charged top-layer sulfur vacancy $V_S^-$ in type 2 Jahn-Teller distorted state, and **b**, the negatively charged carbon substitutional defect $C_S^-$, excluding the influence of spin-orbit coupling (left) and total density of states including spin-orbit coupling (right). The Fermi level in (a, b) was set to fit the energy of the HODO in experiment.

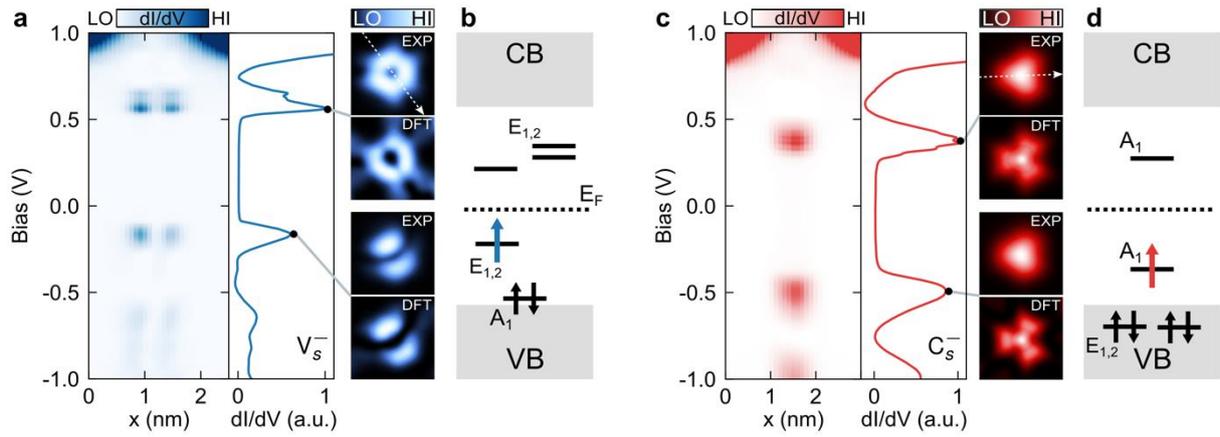

**Supplementary Fig. 3. Electronic structure of spin defects. a**, d$I$/d$V$ line spectroscopy of $V_S^-$ (indicated by the white dotted line in the d$I$/d$V$ orbital map) with integrated d$I$/d$V$ spectrum (middle). dI/dV orbital maps of the states indicated by the two black dots are shown on the right (image size: 1.7 nm × 1.7 nm, +565 mV and -170 mV) along with corresponding DFT-calculations (Supplementary Note 2). **b**, Energy level diagram of $V_S^-$ in-gap states (CB: conduction band, VB: valence band). In the neutral state, the defect orbitals are split into one $A_1$ and two $E_{1,2}$ states with orbital degeneracy. Including spin-orbit coupling, charging, and Jahn-Teller distortion yields a highest occupied defect orbital (HODO) along with multiple lowest unoccupied defect orbitals (LUDO) and a single unpaired spin in one of the $E_{1,2}$ states (blue arrow). Black arrows indicate the occupied $A_1$ spin states. **c**, d$I$/d$V$ measurements of $C_S^-$ (d$I$/d$V$ map image size: 1.7 nm × 1.7 nm; +400 mV and -480 mV). The $C_S^-$ defect shows nearly identical orbital shapes for the occupied (-480 meV) and unoccupied (+400 meV) states, which indicates a single open-shell defect configuration. Right: d$I$/d$V$ orbital maps are in excellent agreement with DFT calculations (see also Supplementary Note 2) for both orbital symmetry and energetic alignment. **d**, Corresponding energy level diagram: The $C_S^-$ hosts a single spin in the half-occupied $A_1$ orbital. [Black arrows indicate occupied $E_{1,2}$ states. Setpoints: (a, c) $V_{DC} = -1.5$ V, $I = 50$ pA, $V_{mod} = 10$ mV].

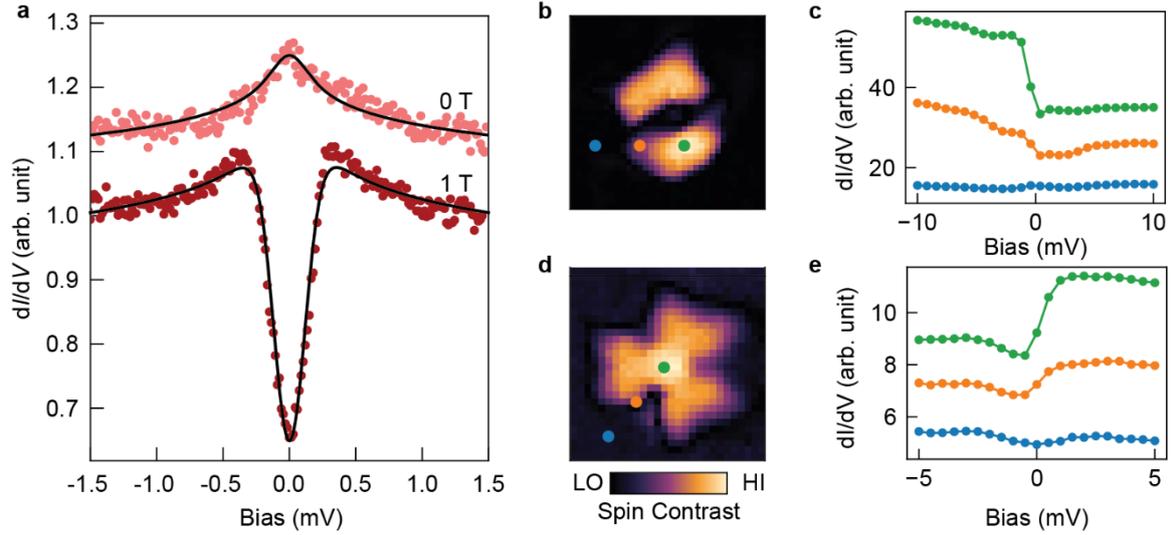

**Supplementary Fig. 4. Additional d$I$/d$V$ spectroscopy and spin density maps. a,** d$I$/d$V$ spectra acquired on a carbon substitution ($C_S^-$) at zero magnetic field and at an out-of-plane field of 1 T. At zero field, the spectrum exhibits a weak zero-bias Kondo resonance, which splits under applied magnetic field. The spectra are vertically offset for clarity. Black lines are fits based on inelastic electron spin transport calculations of a spin-½ system with g = 2 [38] (Setpoints: a, $I$ = 100 pA, $V_{DC}$ = 5 mV, $V_{mod}$ = 50 µV) **b-e,** Spin contrast maps of **b**, a sulfur vacancy ($V_S^-$) and **d**, a carbon substitution ($C_S^-$), acquired with a spin-polarized STM tip. In the presence of a localized spin, inelastic electron tunneling leads to a pronounced asymmetry in the differential conductance at opposite bias polarities near zero bias. [38,39]. The maps are calculated from $|dI/dV\,(+V_{asym}) - dI/dV\,(-V_{asym})|$ with $V_{asym}$ = 10 mV (5 mV) for $V_S^-$ ($C_S^-$). **(c, e)** d$I$/d$V$ point spectra recorded at the positions indicated in (b, d), showing enhanced conductance asymmetry at negative bias in regions of high spin contrast and negligible asymmetry in regions with low contrast. All spectra were acquired at zero magnetic field and are vertically offset for clarity. [Imaging parameters: 1.12 nm × 1.12 nm; (b-c) $V$ = 10 mV, $I$ = 200 pA, $V_{mod}$ = 0.8 mV; (d-e) $V$ = 5 mV, $I$ = 500 pA, $V_{mod}$ = 0.5 mV].

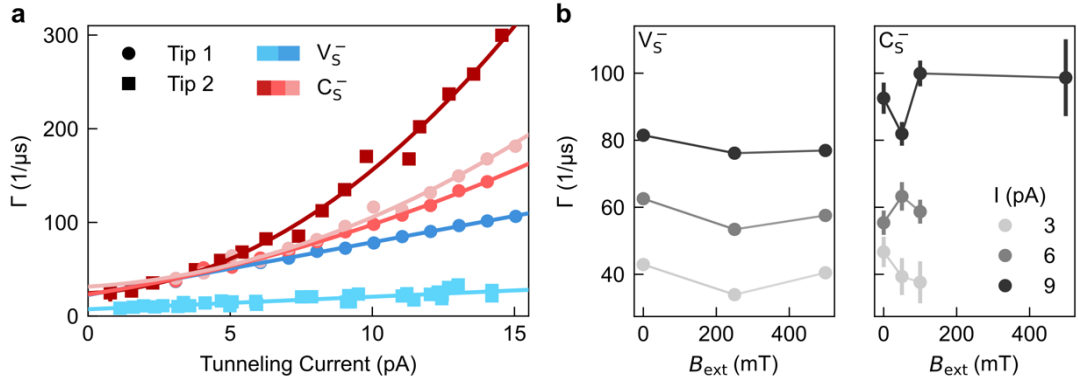

**Supplementary Fig. 5. Current- and magnetic field-dependent SRS measurements. a**, The total transition rate $\Gamma$ as a function of tunneling current $I$ for both defect types $V_S^-$ (blue) and $C_S^-$ (red) for two different tips and different defects ($V_{AC}$ = 2 mV, $|V_{DC}|$ = 1.5 − 2 mV, $B_{ext}$ = 0 T). In both cases, $\Gamma$ increases with increasing $I$, indicating that inelastic scattering of tunneling electrons induces spin flips. $V_S^-$ exhibits both a lower zero-current relaxation rate and a more linear dependence on tunneling current than $C_S^-$, consistent with reduced local coupling to tunnelling electrons due to its spatially distributed (two-lobe) orbital overlap (HODO shown in Supplementary Fig. 3a) with the tip. From the slope in the linear part, we find that the inelastic scattering probability is $P_{T_1}^{V_S^-} = \left(\frac{\Delta\Gamma}{\Delta I}\right) \cdot e \approx (45 \pm 1)\%$ per electron. **b**, Transition rate $\Gamma$ for different external magnetic field $B_{ext}$. Data taken for different tunnelling currents $I$ and both defect types $V_S^-$ (left), $C_S^-$ (right) ($V_{AC}$ = 2 mV, $|V_{DC}|$ = 1.5 mV for $V_S^-$ and $|V_{DC}|$ = 2 mV for $C_S^-$). In general, $\Gamma$ does not show a strong dependence on magnetic field.

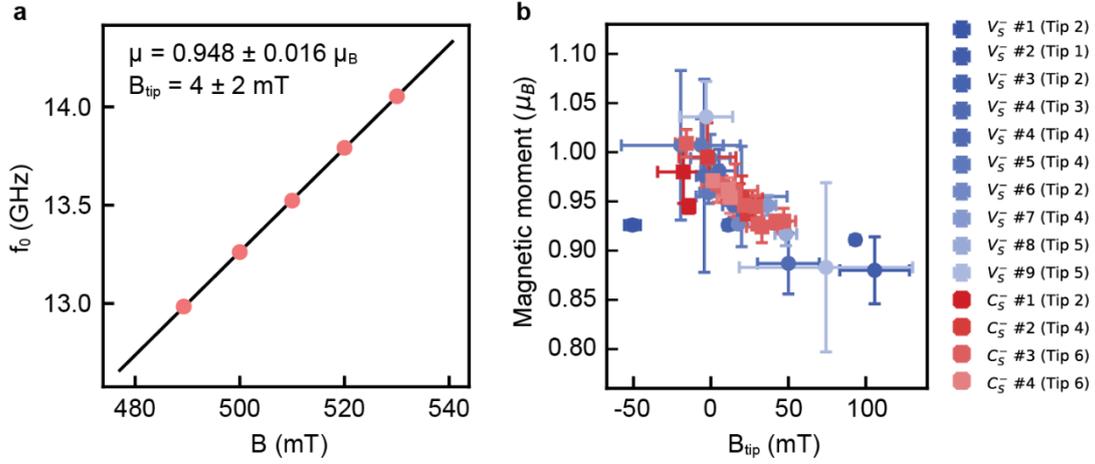

**Supplementary Fig. 6. Effective magnetic moments extracted from ESR measurements.** **a**, ESR resonance frequency $f_0$ of $C_S^-$ as a function of magnetic field, extracted from Fig. 2d. A linear fit using $hf_0=2\mu(B+B_{tip})$ yields an effective magnetic moment of $\mu_{C_S^-} = 0.948 \pm 0.016\ \mu_B$ and an effective magnetic tip field of $B_{tip} = 4 \pm 2$ mT (Setpoints: $I = 5$ pA, $V_{DC} = 20$ mV, $V_{RF} = 10$ mV). **b**, Effective magnetic moment $\mu$ extracted from linear fits of the ESR resonance frequency as a function of $B$ for sulfur vacancies ($V_S^-$, blue) and carbon substitutions ($C_S^-$, red). They are plotted as a function of the effective magnetic tip field $B_{tip}$, also obtained from the fits. Each data point corresponds to a different defect-tip combination ($V_S^-$: 9 defects measured with 6 tips; $C_S^-$: 4 defects measured with 3 tips). Horizontal and vertical error bars are derived from the linear fits. The extracted magnetic moment shows a dependence on $|B_{tip}|$, reflecting the influence of the tip field on the resonance condition $f_0 \sim B$ as pointed out by Ref. [57]: An apparent change in $\mu$ occurs, if $B_{tip}$ also changes with $B$ and is minimized for vanishing $B_{tip}$. Thus, we only average values obtained for $|B_{tip}|<10$ mT, which yields $\mu_{V_S^-} = (0.992 \pm 0.011)\ \mu_B$ and $\mu_{C_S^-} = (0.975 \pm 0.007)\ \mu_B$. In addition to the influence of $B_{tip}$, variations in $\mu$ may also arise from differences in the local environment of the spin center, such as local strain or Jahn-Teller distortion.

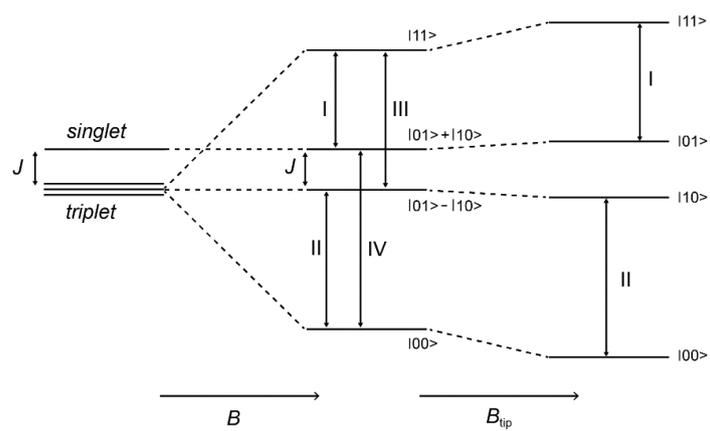

**Supplementary Fig. 7.** Schematic energy level diagram of the two spin 1/2 as a function of $B$ and $B_{\text{tip}}$ for given $J$ (ferromagnetic). Labels I, II, III, IV denote the possible transitions when performing ESR on the first spin.

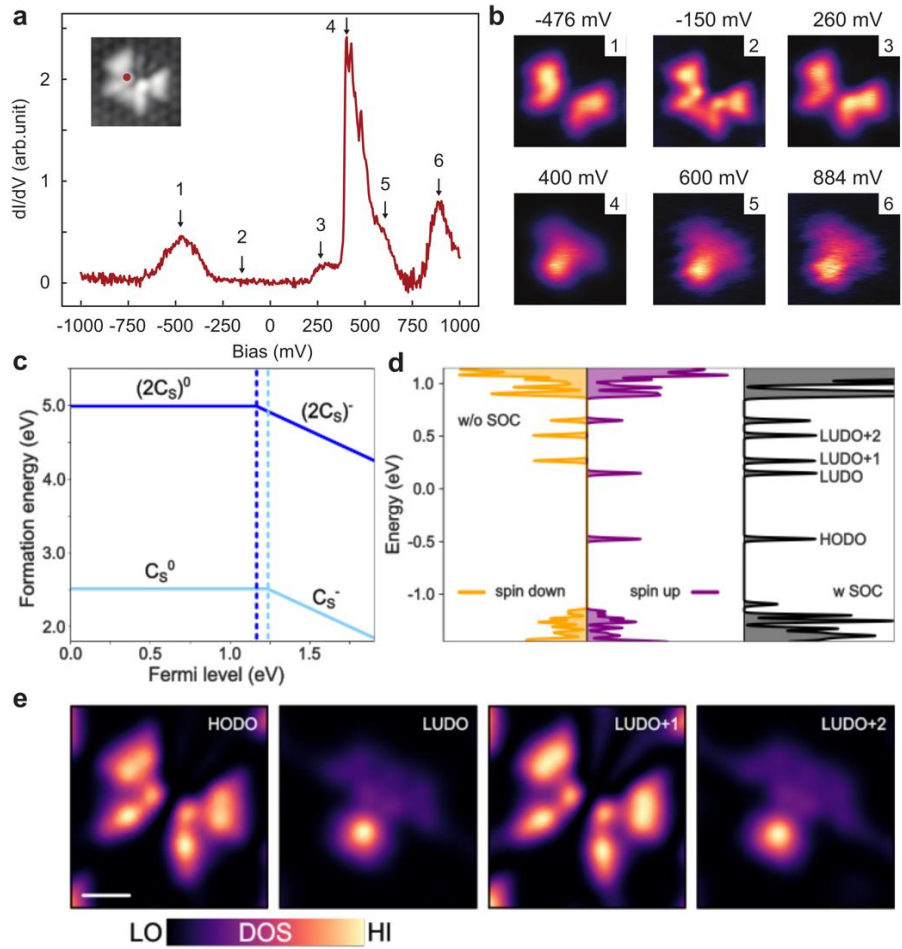

**Supplementary Fig. 8. Experimental and DFT-calculated electronic structure of the carbon disubstitution $(2C_S)^-$ defect. a**, d$I$/d$V$ spectrum measured on the closest possible (0,1) configuration of two carbon substitution defects, forming a carbon dimer defect denoted $(2C_S)^-$ (Fig. 4c). Compared to the isolated $C_S^-$ defect (Supplementary Fig. 3c), the spectrum exhibits additional resonances and pronounced changes in spectral weight, indicating the emergence of new electronic states. Arrows mark the bias voltages at which spatially resolved d$I$/d$V$ maps were acquired. The inset shows the STM topography of the defect, with the position of the spectroscopy measurement indicated (setpoint: $I = 100$ pA, $V_{DC} = 1$ V). **b**, Corresponding d$I$/d$V$ maps (1.92 nm × 1.92 nm) recorded at the bias voltages indicated in (a), revealing the spatial evolution and symmetry of the defect's electronic states. The maps are distinct from those of isolated carbon substitution defects, reflecting the modified electronic structure of the $(2C_S)^-$ complex. In addition to the resonant states marked in (a), the in-gap state at –150 mV is shown. **c**, DFT-calculated formation energies of a single carbon substitution defect ($C_S$, light blue) and a carbon substitution dimer ($2C_S$, dark blue) in nearest-neighbor configurations as a function of Fermi level position. A transition between a charged and neutral defect can be identified by the change of slope, as the Gibbs free energy of defect formation changes linearly with Fermi level when the defect is charged by an amount $q$ [58]. Here $q$ changes from zero (neutral) to $-e$, with $e$ being the elementary charge. **d**, Density of states of $(2C_S)^-$ calculated without spin–orbit coupling (left, spin-polarized) and including spin–orbit coupling (right). The Fermi level was set to fit the energy of the highest occupied defect orbital (HODO)

in experiment. **e**, Simulated images of four in-gap defect states of $(2C_S)^-$ calculated by DFT (Scale bar: 0.5 nm).

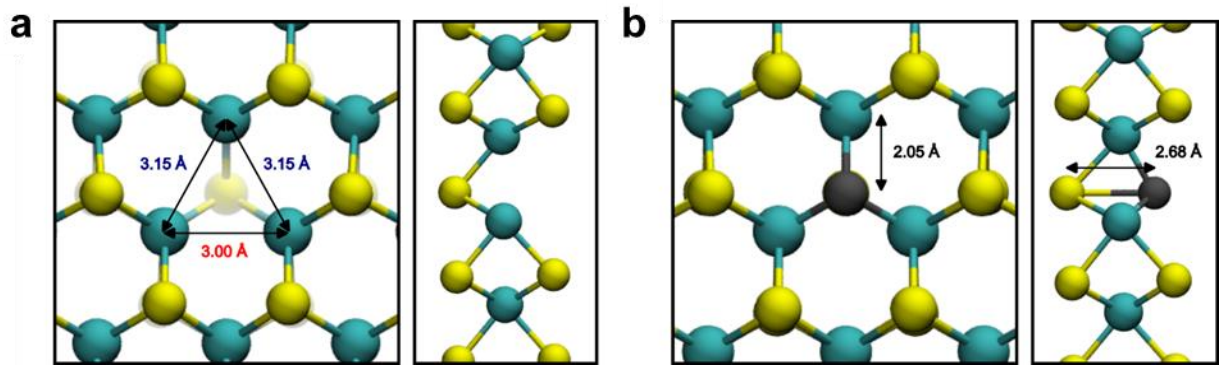

**Supplementary Fig. 9. Atomic structure model of point defects computed by density functional theory. a,** Top (left) and side view (right) of a single negatively charged top-layer sulfur vacancy $V_S^-$ in the type 2 Jahn-Teller distorted state. The arrows mark the distances between Mo atoms surrounding the sulfur vacancy. **b,** Top (left) and side view (right) of a substitutional carbon atom in the negative charge state $C_S^-$. The arrows mark respective bond lengths.

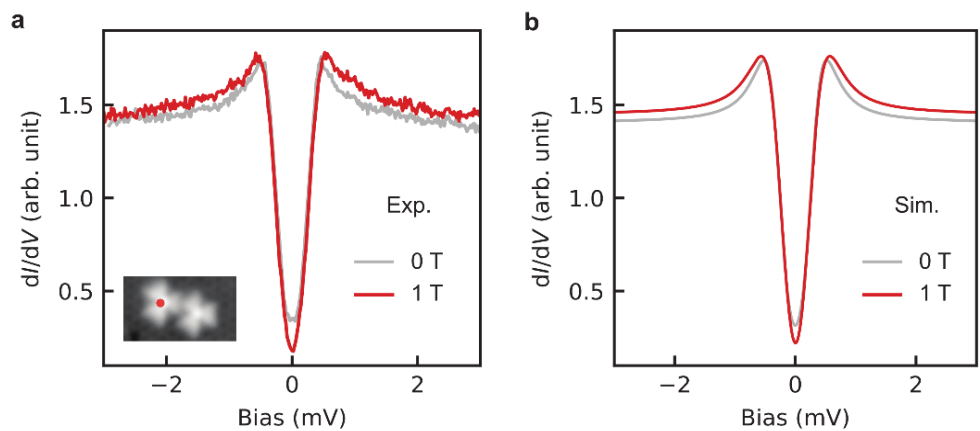

**Supplementary Fig. 10. Comparison between experiment and simulation of the differential conductance of the $C_S^-$ defect dimer**. **a,** Experimental d$I$/d$V$ spectra measured on the left $C_S^-$, as indicated in the inset, at 0 T (gray) and at 1 T (red, Setpoints: $V_{DC} = 10$ mV, $I = 300$ pA, $V_{mod} = 50$ μV). **b,** Corresponding simulated d$I$/d$V$ spectra at 0 T (gray) and 1 T (red). The simulations are performed using inelastic electron spin-transport calculations for two antiferromagnetically coupled spins $S = 1/2$, with an exchange coupling $J = 354$ μeV and g = 2 for both spins.

**Supplementary Note 1. Stochastic Resonance Spectroscopy Analysis**

In this section, we discuss in greater detail the results of stochastic resonance spectroscopy (SRS). We point to Ref. [41] for an in-depth discussion on the SRS technique. Moreover, we follow the framework of spin relaxation established in Ref. [42] for inelastic scattering between on-surface spins and tunneling electrons. We employ SRS to quantify the spin relaxation time $T_1$, identify dominant relaxation channels, and probe the spatial extent of individual defect spins. In STM-based SRS, an oscillating bias voltage $V_{AC}$ drives stochastic transitions between the two spin states $|0\rangle$ and $|1\rangle$, and the resulting frequency-dependent tunneling current yields the transition rate $\Gamma = \Gamma_{0\to1} + \Gamma_{1\to0}$, with $T_1 = 1/\Gamma_{1\to0} \approx 1/\Gamma$ (Fig. 2c) [41]. This is observed as a characteristic step at $f_{AC} = \Gamma/2\pi$.

We first investigate the dependence of $\Gamma$ on tunneling current $I$ in greater detail. For both defect types and different tips, the transition rate increases strongly with current (Supplementary Fig. 5a), indicating that relaxation is dominated by scattering processes involving tunneling electrons and nearby metallic electrodes. While $\Gamma(I)$ is linear for $V_S^-$ defects over the measured range, $C_S^-$ defects exhibit an additional non-linear contribution, consistent with enhanced coupling to the tip electrode (see below).

To model this behavior, we write $\Gamma$ as the sum of inelastic tunneling processes $\Gamma_{ij}$, in which electrons from the tip (t) and sample (s) exchange energy with the spin [42] [assuming $V = 0$ V]:

$$\Gamma = \Gamma_{t-t} + \Gamma_{t-s} + \Gamma_{s-t} + \Gamma_{s-s} = \frac{P_{T_1} E_{01}}{e^2} ( G_{t-t} + G_{t-s} + G_{s-t} + G_{s-s}) \quad (S1)$$

Here, $P_{T_1}$ denotes the probability that a tunneling electron induces a spin relaxation event (assumed to be identical for all electrode pairs) and $E_{01}$ is the energy splitting between the ground and excited spin states. The conductances $G_{t-s} = G_{s-t} = I/V$ describing tunneling from tip to substrate and vice versa are equal for a symmetric junction. The terms $G_{s-s}$ and $G_{t-t}$ account for scattering processes in which electrons originate from and return to the same electrode. For an additionally applied bias voltage $V$ and assuming $k_B T \ll E_{01} \leq |eV|$ the transition rate can be rewritten as [Eq. (S32) and (S33) in Ref. [42]]

$$\Gamma \approx \Gamma_0 + 2\frac{P_{T_1}}{e}|V|G_{t-s} + \Gamma_{t-t}, \quad (S2)$$

where $\Gamma_0$ denotes the transition rate in the absence of tunneling current corresponding to the intrinsic lifetime $T_{1,0} \approx 1/\Gamma_0$. $\Gamma_{t-t}$ denotes scattering with the tip electrode only, responsible for a quadratic dependence (see below).

The fits to Eq. (S2) [and using (S3), see below] shown in Supplementary Fig. 5a (see Supplementary Table 1) yield relaxation probabilities $P_{T_1}$ in the range of 10–45%. These values are comparable to those reported for other spin-1/2 molecular systems, such as FePc on MgO/Ag(001) ($P_{T_1} = 32\%$) [29] and are consistent with expectations from inelastic electron spin-transport theory [38]. In contrast, it can be several orders of magnitude smaller in the presence of magnetic anisotropy, e.g. Fe on MgO/Ag(001) (S = 2, $P_{T_1}$ = 0.0003% ) [42]. Variations between defects and tips are attributed to differences in tip spin polarization and local geometry of tip and defect. The extracted intrinsic lifetimes are consistent across datasets, yielding an average $\langle T_{1,0} \rangle$ = 40 ± 5 ns, with one longer-lived outlier attributed to a double-tip configuration.

Consistent with the analysis of $P_{T_1}$, we attribute the current-independent relaxation rate $\Gamma_0$ primarily to inelastic scattering with substrate electrons. This mechanism has been identified as the dominant relaxation channel in closely related systems, such as Fe adatoms on MgO/Ag(001) [42]. Applying the same considerations here, point contact measurements on $V_S^-$ (in which the tip is approached atop the MoS2 defect until current saturation is reached) yield a substrate conductance of $G_S \approx 2.8\ \mu S$. Using this value, we can estimate an associated substrate-scattering-limited lifetime $T_{1,s-s} = \frac{e^2}{k_B T} \frac{G_0}{G_S^2} \approx 370$ ns [See analysis in Supplementary Section 7 in Ref. [27]]. Because the experimental temperature ($T$ = 50 mK) represents a lower bound on the effective electron temperature, this estimate constitutes an upper bound and supports that the residual relaxation mechanism is dominated by substrate-induced electronic scattering.

$\Gamma(I)$ is linear for $V_S^-$ defects, while $C_S^-$ defects exhibit an emergent quadratic dependence. We attribute this to the proximity to the tip electrode which increases $G_{t-t}$. For a symmetric tunneling barrier, $G_{t-t} = G_{t-s}^2/G_{s-s}$ and we can substitute in Eq. (S2):

$$\Gamma_{t-t} = \frac{P_{T_1} E_{01}}{e^2} \frac{G_{t-s}^2}{G_{s-s}} \approx \Gamma_0 (\frac{G_{t-s}}{G_{s-s}})^2 \tag{S3}$$

using $\Gamma_0 \approx \Gamma_{s-s}$. Here, $G_{s-s}$ should be understood as an effective substrate-substrate conductance within the scattering model [42]: It captures all substrate-related contributions to spin relaxation, such as $P_{T_1}$, $E_{01}$ and the number of conductance channels. Assuming a single conduction channel, this yields an effective substrate conductance $\langle G_S \rangle \approx \sqrt{\langle G_{s-s} \rangle G_0} \approx$ 520 ± 80 nS for $C_S^-$. This can be compared to independent point-contact measurements on $C_S^-$ defects, which estimate $G_S$ to be 30 nS − 4.2 μS. These measurements are experimentally more challenging than for adatoms on surfaces, since the tip apex is often modified before a saturation current is reached. Despite the large variations, it agrees with the results from the rate equation model for $\langle G_S \rangle$ and therefore the emergence of a quadratic term for $C_S^-$ is reasonable. We attribute the absence of the latter for $V_S^-$ to the spatial delocalization of the $V_S^-$ spin density, which reduces coupling to the tip electrode.

Additionally, we examine the dependence of $\Gamma$ on an external out-of-plane magnetic field $B_{\text{ext}}$. As shown in Supplementary Fig. 5b, $\Gamma$ remains strongly dependent on tunneling current, while its variation with $B_{\text{ext}}$ is limited over the range 0–500 mT and shows no systematic trend. This variation is small compared to the much stronger dependence on current or tip position. We therefore attribute the residual field dependence of $\Gamma$ to changes in the tip spin polarization

$P(B_{ext})$ and the effective tip field $B_{tip}(B_{ext})$, rather than to an intrinsic magnetic-field dependence of the defect spin.

| Defect (Tip) | $V$ (mV) | $P_{T_1}$ (%) | $\Gamma_0$ (1/μs) | $T_{1,0}$ (ns) | $G_{s-s}$ (nS) |
|---|---|---|---|---|---|
| $V_S^-$ (#2) | −1.5 | 45 ± 1 | 23 ± 1 | 44 ± 1 | – |
| $V_S^-$ (#1) | −2.0 | 11 ± 1 | 7 ± 1 | 137 ± 6 | – |
| $C_S^-$ (#1) | +2.0 | 12 ± 9 | 25 ± 2 | 41 ± 4 | 2.3 ± 0.1 |
| $C_S^-$ (#2) | +2.0 | 37 ± 7 | 23 ± 2 | 43 ± 6 | 4.5 ± 0.2 |
| $C_S^-$ (#2) | +2.0 | 15 ± 18 | 31 ± 8 | 32 ± 8 | 3.8 ± 0.2 |

**Supplementary Table 1.** Fit parameters obtained from modeling the SRS transition rate $\Gamma(I)$ using Eq. (S2) for different defect types and tips. Since $\Gamma(I)$ is linear for $V_S^-$ defects, no values are obtained for $G_{s-s}$.

**Supplementary Note 2. Density functional theory calculations for $V_S^-$, $C_S^-$ and $(2C_S)^-$**

DFT calculations were carried out to validate the nature of point defects $V_S^-$ and $C_S^-$ studied in the experiments. We show the relaxed atomic structure models for $V_S^-$ in Supplementary Fig. 9a and $C_S^-$ in Supplementary Fig. 9b as resulting from DFT calculations. For $V_S^-$, the surrounding lattice is distorted due to a Jahn-Teller effect. In general, two different types of Jahn-Teller distortions exist [24]. To match the experimental measurements, a Jahn-Teller distortion of type 2 as defined in Ref. [24] is depicted. In this configuration there are two longer bonds of 3.15 Å and one shorter bond of 3.0 Å between the Mo atoms surrounding the sulfur vacancy. For the $C_S^-$ defect, all three bonds between the substitutional C and the surrounding Mo atoms are of equivalent length (2.05 Å).

**Sulfur vacancy ($V_S^-$):** Supplementary Fig. 2a shows the electronic properties of the negatively charged sulfur vacancy $V_S^-$ in the Jahn-Teller distorted state (type 2). Here, the density of states (DOS) is plotted in a spin-polarized form, without spin-orbit coupling (SOC) on the left and in non-collinear form including SOC on the right. The Fermi-level position is defined such, that the energy of the highest occupied defect orbital (HODO) matches the experimentally determined value of the sulfur vacancy shown in Supplementary Fig. 3a. From the plot on the left of Supplementary Fig. 2a it becomes apparent that all in-gap states of $V_S^-$ are spin-polarized. PBE calculations (not shown) underestimate the HODO-LUDO-splitting with 0.3 eV compared to the experimental value of 0.735 eV. Introducing Heyd-Scuseria-Ernzerhof (HSE) hybrid functionals and a mixing parameter of 10% (same as for $C_S^-$) yields a splitting of 0.62 eV which is in good agreement with experiment and shown in Supplementary Fig. 2a. In Supplementary Fig. 3a, a comparison between simulated STM images and experimentally measured d$I$/d$V$ maps for the in-gap states are presented. The HODO is shown separately on the left, as it is well-isolated from other states. For the lowest unoccupied defect orbitals (LUDOs) a combination of the three states LUDO, LUDO+1 and LUDO+2 is shown on the right as these states are all close in energy and cannot easily be disentangled. The simulated images were generated by evaluating the defect DOS at a distance of 4.3 Å from the top-layer sulfur plane.

We can rationalize the DFT results by the interaction of the sulfur vacancy with the surrounding MoS$_2$ lattice: A single sulfur vacancy in the neutral charge state $V_S^0$ in monolayer MoS$_2$ leaves three neighboring Mo atoms in a C$_{3v}$ environment; their three dangling-bond orbitals form symmetry-adapted linear combinations (SALC) that split into one A$_1$ singlet and two E pairs with orbital degeneracy. In the neutral case all E orbitals are unoccupied. Spin-orbit coupling splits them into two Kramers doublets (often labelled E$_{5/2}$ and E$_{3/2}$) [22]. In the negatively charged state $V_S^-$ the configuration is A$_1^2$E$^1$ (S = ½). In this case, we do not find any significant shifts or splitting when including SOC, see Supplementary Fig. 2a. Therefore, we conclude that the unpaired electron behaves as a spin-1/2 system with a g-factor close to 2.

**Carbon substitution ($C_S^-$):** In Supplementary Fig. 2b the electronic properties of $C_S^-$ are shown. Including the density of states. On the left the spin-polarized DOS calculated without the influence of SOC is depicted. As can be seen, both the HODO as well as the LUDO are spin-polarized. On the right side the full spin-averaged DOS including SOC is shown. The Fermi level is again shifted such that the energy of the HODO matches the experimentally determined value of the carbon substitution in Supplementary Fig. 3c. The DFT calculations using the PBE functional yield an energy separation of 0.57 eV between HODO and LUDO underestimating the actual separation which was experimentally determined to be 0.88 eV. As in the case of sulfur vacancies, a reasonable match with experiment can be obtained when HSE hybrid functionals are employed with a mixing parameter of 10%, which is plotted in Supplementary Fig. 2b.

The spin state can thus be rationalized as in the following: Substituting carbon at a sulfur site (C$_S$) hybridizes the carbon's 2p$_z$-orbitals with the three Mo SALCs to form bonding and antibonding A$_1$ and E combinations with lower lying E orbitals [17]. In the neutral case ($C_S^0$) both E orbitals are filled by two electrons, as carbon has two fewer valence electrons compared to sulfur. The A$_1$ orbital remains empty. In the charged state $C_S^-$ the configuration is E$^4$ A$_1^1$ (S = 1/2). Alternatively, one could also rationalize this in the double acceptor picture: Here, removing two protons from S-site leads to S-p$_z$ state lifting up from the valence band. In either case, the extra electron added to the A$_1$ orbital splits the singly and doubly occupied states by Coulomb repulsion [54]. Hence, for $C_S^-$ a single spin S = ½ resides in the A$_1$ level with an almost spin-only, near-isotropic g-factor. No Jahn-Teller effect is present as A$_1$ has no orbital degeneracy. The defect therefore maintains its three-fold symmetry.

**Carbon Di-Substitutions (2C$_S$)$^-$:** Supplementary Fig. 8c shows the formation energy of an isolated carbon substitution defect C$_S$ and of carbon substitution dimers (2C$_S$) as a function of the Fermi level. Within the MoS$_2$ band gap, both the isolated C$_S$ and the (2C$_S$) dimer are stable only in the neutral, $C_S^0$ and (2C$_S$)$^0$, as well as singly negatively charged states, $C_S^-$ and (2C$_S$)$^-$. Notably, the dimer does not support a doubly charged state in the given range and consequently only one electron can be accommodated at most. We attribute this to the enhanced Coulomb repulsion arising from the close proximity of the two defects. In the formation energy calculations, the sulfur chemical potential is chosen in the Mo-rich limit [$\mu_{MoS2} - 2\mu_S$], while the carbon chemical potential is referenced to graphene. Finite-size and charge-correction effects are treated following Ref. [58].

The electronic structure of the $(2C_S)^-$ is shown in Supplementary Fig. 8d. The spin-polarized density of states calculated without spin–orbit coupling (SOC) reveals a singly occupied in-gap state (left) as well as unoccupied in-gap states (LUDOs) above the Fermi level, and the inclusion of SOC does not qualitatively alter the electronic structure (right). As for $V_S^-$ and $C_S^-$ defects, HSE hybrid functionals with a 10% mixing parameter were employed in the DFT calculations. Supplementary Fig. 8e shows simulated STM images of all HODO and LUDO states.

The experimental $dI/dV$ maps of $(2C_S)^-$ states are shown in Supplementary Fig. 8a. Overall, we find close agreement with DFT calculations: The experimental maps at –476 mV, +400 mV, +260 mV, and +884 mV match the calculated HODO, LUDO, LUDO+1 and LUDO+2 states, respectively. The order of LUDO and LUDO+1 differs between DFT and experiment, likely due to their small splitting in energy (~150 meV), which is difficult to capture accurately via DFT wavefunctions. Only the maps at –476 mV and +400 mV are shown in the main text, as they correspond to the most dominant spectral features. Overall, the DFT calculations reproduce the correct number of in-gap states and their spatial structure with good agreement.

**Supplementary Note 3. Spin–spin interaction in engineered carbon defect dimers**

**Magnetic coupling between two spins:** In this section, we discuss in greater detail the spin-spin-coupling between two $C_S^-$ defects. We closely follow the discussions in reference [39,59-61], where the respective spin Hamiltonians are introduced and further discussed. In the (2,2) configuration shown in Fig. 4a, two $C_S^-$ spins form a weakly coupled dimer, which we infer by the appearance of two distinct ESR resonances. This behavior is well described by an effective Hamiltonian for two coupled spin-½ [39,59-61]

$$H = g\mu_B(B_z + B_{\text{tip}})S_{1,z} + g\mu_B B_z S_{2,z} + J\,\mathbf{S}_1 \cdot \mathbf{S}_2 \quad \text{(S4)}$$

where $B_z$ is the external magnetic field, $B_{\text{tip}}$ is the effective local magnetic field generated by the STM tip acting on spin 1, and $J$ is the Heisenberg exchange coupling. The magnetic dipolar interaction is neglected in this discussion ($D \approx 38$ MHz for two spins with g = 1.95 separated by 1.09 nm), since it is one order of magnitude smaller than $J$.

The Hamiltonian yields two Zeeman product states $|00\rangle$ and $|11\rangle$, and two states $|-\rangle$ and $|+\rangle$ that are superpositions of $|01\rangle$ and $|10\rangle$, mixed via the exchange interaction. Their eigenenergies are given by[59]:

$$E_{00} = -\tfrac{1}{2}g_z\mu_B(2B + B_{\text{tip}}) + \tfrac{1}{4}J \quad \text{(S5)}$$

$$E_- = -\tfrac{1}{2}\sqrt{\delta^2 + J^2} - \tfrac{1}{4}J \quad \text{(S6)}$$

$$E_+ = \frac{1}{2}\sqrt{\delta^2 + J^2} - \frac{1}{4}J \qquad (S7)$$

$$E_{11} = \frac{1}{2}g_z\mu_B(2B + B_{tip}) + \frac{1}{4}J \qquad (S8)$$

Where $\delta = g\mu_B B_{tip}$. A schematic energy level diagram is shown in Supplementary Fig. 7.

Since the exchange coupling is much smaller than the Zeeman energy, four ESR transitions are allowed in principle. In the experiment, however, only two resonances are observed (Fig. 4a), because $B_{tip}$ tunes the mixed states into Zeeman product states (Supplementary Fig. 7). From Eqs. (S5)-(S8), the difference between the two resonance frequencies directly gives the magnetic exchange coupling between the two spins, $\Delta f = (E_{11} - E_+) - (E_- - E_{00}) = J \approx -400$ MHz (~2 µeV). Moreover, the higher intensity of the resonance at higher frequencies, reflecting the thermal occupation of the initial states, is consistent with ferromagnetic exchange[39,59].

The exchange coupling can be tuned by varying the distance and relative orientation of the two defects. In Fig. 4b, the two spins arrange in a (1,3) configuration, with a separation of 1.136 nm, which is slightly larger than that of the (2,2) configuration shown in Fig. 4a. Nevertheless, the real-space orbital overlap is enhanced due to the relative alignment of the defect orbitals, resulting in a substantially stronger exchange interaction. In this configuration, the coupling is antiferromagnetic. This change in sign and magnitude of $J$ is directly reflected in the spectroscopic signatures. The experimental d$I$/d$V$ spectra acquired at zero field and at 1 T (Supplementary Fig. 10a) are well reproduced by inelastic electron spin-transport simulations assuming two antiferromagnetically coupled spin-1/2 ( $1\,\mu_B$ ) with $J = 354\,\mu$eV (Supplementary Fig. 10b) [38]. Simulations assuming ferromagnetic coupling fail to reproduce the observed inelastic step positions and intensities, confirming the antiferromagnetic nature of the interaction in this geometry.